\documentclass[a4paper,10pt]{article}
\usepackage[latin1]{inputenc}
\usepackage{amsmath}
\usepackage{graphicx}
\usepackage{epsfig}

\usepackage{amsfonts}
\begin{document}

\begin{center}
{\Large\bf Tracking quintessence: a dynamical systems study}
\\[15mm]
Nandan Roy \footnote{E-mail: nandan@iiserkol.ac.in} and
Narayan Banerjee \footnote{E-mail: narayan@iiserkol.ac.in}

{\em Department of Physical Sciences,~~\\Indian Institute of Science and Educational Research-Kolkata,~~\\Mohanpur Campus, West Bengal 741252, India.}\\[15mm]
\end{center}

\begin{abstract}
With the tracking condition, the stability of quintessence solutions are examined. It is found that there is only one physically relevant fixed point for the system generically. Two specific examples of quintessence potentials are worked out in the frame work.
\end{abstract}


\section{Introduction:}
Notwithstanding the universal attractive nature of gravity, the universe appears to be going through an accelerated phase of expansion. This strange behaviour has strong observational evidence\cite{r1} and is a widely accepted reality. The acceleration is reported to be a recent affair, started well within the matter dominated regime\cite{r20}. That this acceleration has set in after a long stint of decelerated exapnsion is a theoretical requirement as well\cite{r2}. But the matter that drives this acceleration is still a mystery, which is neither detected observationally nor has any single firmly accepted theoretical model. A rejuvenated introduction of the  cosmological constant $\Lambda$ does very well in explaining this recent acceleration, but $\Lambda$ has its own problems\cite{r3}. Amongst a host of alternatives, a quintessence field\cite{r4} attracts a lot of attention. A scalar field, minimally coupled to gravity, endowed with a potential, is called a quintessence fi
 eld. The idea is that the potential may give an effective negative pressure which would result in providing the required anti-gravity effect and thus drive an accelerated expansion. However there is no indication of a clear verdict in favour of a potential in terms of a theoretical basis. 

Along with the problem of finding a suitable driver (generally called a dark energy) of the present acceleration, the associated problem that crops up is that why at the present epoch this dark energy bears a constant ratio, of order unity, to the dark matter. The cosmological constant, if that is the solution to the dark energy problem, should be extremely fine tuned to the density of matter or radiation at the very early stage\cite{r5}.\\

Thus there are two problems, one is the absence of a sound theoretical basis in favour of any particular quintessence potential, and the other is that of a fine tuning of initial values. There is no serious remedy for the first one. Attempts have been made to model the acceleration with an arbitrary potential using two scalar fields\cite{r}.\\

One way to address the second problem is to look for a quintessence field which `tracks' the matter density, i.e. evolves at almost the same rate but below the level of the dark matter but slowly catches up so as to eventually lead the scenario only at a later stage \cite{r5, r6}.\\

Naturally there has been a lot of work in this direction, where the quintessence field is a tracker. Johri \cite{r7} looked for quintessence potentials which are trackers. Urena-Lopez etal \cite{r8} gave a tracker solution which acts as a quintessence. Sahlen, Liddle and Parkinson reconstructed a quintessence potential and checked its tracking viability \cite{r9}. In order to resolve the coincidence problem, Dodelson, Kaplinghat and Stewart \cite{r10} invoked an oscillating potential as the quintessence field. Wang, Chen and Chen \cite{r11} discussed the performance of some tracker field quintessence potentials against observational data. They found that tracker potentials of the form $exp(\frac{M_p}{\phi})$ or [$exp(\frac{\gamma M_p}{\phi}) - 1 $] are clearly unsuitable.\\

The primary motivation of the present work is to find the stability of the solutions for FRW cosmological models with a pressureless fluid and a tracking quintessence field. The field equations are written as an autonomous system, and the fixed points are located. The fixed points which correspond to the physical requirements of the evolution of the universe are picked up. Actually only one generic fixed point could be obtained, which depends on the equation of state parameter $\gamma$ for the fluid and the fractional rate of decrease of the quintessence potential. Attractor solutions in scalar field cosmology and some of their applications in an inflationary scenario and also in the quintessence scenario have also been discussed by Ng, Nunes and Rosati \cite{r12}. \\

Albeit its being not so extensively used, a dynamical systems analysis is not new in cosmology. We refer to the volume edited by Ellis and Wainwright and work of Coley for reviews \cite{r13}. Lara and Castagnino \cite{r14} presented a dynamical systems analysis for an FRW universe with a number of non minimally coupled scalar fields. Gunzig et al \cite{r15} discussed a spatially flat FRW universe with a scalar field. The motivation was to to look at the nature of the scalar field solutions leading to an inflation at an early epoch. Carot and Collinge also discussed inflationary scalar field cosmologies in the frame work of dynamical systems \cite{r16}. A scalar field leading to a phantom behaviour had been studied by Urena-Lopez \cite{r17}. In the framework of dynamical systems, Brans-Dicke theory and also its equivalnce with  a minimally coupled scalar field have also been discussed\cite{r17}.\\

In section 2, the Einstein field equations for a spatially flat FRW Universe with a scalar field and a pressureless fluid are written as an autonomous system of equations and the relevant fixed points are found out. Section 3 deals with the stability of the solution. Section 4 deals with two specific examples of scalar potentials which can actually drive the present acceleration starting from a decelerated phase. A discussion of the results obtained is given in section 5.

\section{The dynamical system and the fixed points:}
For a spatially flat Robertson Walker Universe, given by the metric
\begin{equation} \label{eq.metric}
ds^2 = dt^2 - a(t)(dr^2 + r^2 d\Omega ^2),
\end{equation}
filled with a perfect fluid and also a scalar field distribution, Einstein field equations are written as
\begin{equation} \label{H}
H^2= \frac{8 \pi G}{3}(\rho_{B} + \frac{1}{2} \dot{\phi}^2 + V(\phi)),
\end{equation}
and
\begin{equation} \label{eq.dH}
\dot{H}=-\frac{8 \pi G}{2}(\gamma \rho_{B} + \dot{\phi}^2).
\end{equation}
Here $H=\frac{\dot{a}}{a}$ is the Hubble parameter, $\rho_{B}$ is the energy density of the perfect fluid given by the equation of state $p _{B}=(\gamma-1) \rho_{B}$,where $\gamma$ is a constant, $\phi$ is the scalar field and $V(\phi)$ is the scalar potential. The conservation equation for the fluid is
\begin{equation} \label{eq.rho}
\dot{\rho}_{B}= -3 \gamma H \rho_{B}.
\end{equation}
The scalar field equation is given by
\begin{equation} \label{eq.phi}
\ddot{\phi} + 3 H \dot{\phi} = - \frac{dV}{d\phi}.
\end{equation}

Not all these equations are independent, and we choose equations  (~\ref{eq.dH}), (~\ref{eq.rho}), and (~\ref{eq.phi}) as the system of equation to be solved. Equations (2) is then considered as a constraint equation. New dimensionless variables $x$ and $y$ are defined as

\begin{equation} \label{eq.trans}
x^2=\frac{k^2 {\phi^{\prime}}^2}{6}, ~~ y^2 = \frac{k^2 V}{3 H^2} ,
\end{equation}
where a prime denotes a differentiation with respect to $N=ln(\frac{a}{a_0})$ and $k^2=8 \pi G$. The value of $a_0$, the present value of the scale factor, is chosen to be unity. The contribution of the scalar field $\phi$ to the density and pressure can be written respectively as 
\begin{equation} \label{eq.rhop}
\rho_{\phi}= \frac{1}{2} {\dot{\phi}}^2 + V(\phi),~~ p_{\phi}= \frac{1}{2} {\dot{\phi}}^2 - V(\phi).
\end{equation}
They can be formally connected by an equation of state 
$p_{\phi}=(\gamma_{\phi}-1) \rho_{\phi},$
thus the equation of state parameter $\gamma_{\phi}$ for the scalar field can be written as 
\begin{equation} \label{eq.gamma}
\gamma_{\phi}=\frac{\rho_{\phi} + p _{\phi}}{\rho_{\phi}} = \frac{{\dot{\phi}}^2 }{\frac{{\dot{\phi}}^2}{2} + V} = \frac{2 x^2}{x^2 + y^2}.
\end{equation}
The relevant equations, namely (~\ref{eq.dH}), (~\ref{eq.rho}) and (~\ref{eq.phi}) can now be written as a 3-dimensional autonomous system (with N as the argument),
 \begin{equation} \label{eq.x}
 x^{\prime} = -3x + \lambda \sqrt{\frac{3}{2}} y^2 +\frac{3}{2} x [2 x^2 + \gamma (1-x^2-y^2)]~~~,
\end{equation}

\begin{equation} \label{eq.y}
y^{\prime} =- \lambda \sqrt{\frac{3}{2}} xy + \frac{3}{2} y [2 x^2 + \gamma (1-x^2-y^2)] ~~~,
\end{equation}
and 
\begin{equation} \label{eq.lambda}
\lambda^{\prime} = -\sqrt{6} \lambda^2 (\Gamma -1) x ~~~,
\end{equation} \\
where $ \lambda = - \frac{1}{k V} \frac{dV}{d \phi}$ and $\Gamma =  V\frac{d^2 V}{d \phi^2} / (\frac{dV}{d \phi})^2$. $\Gamma$ is called the tracker parameter. For tracking, one has to set $\Gamma \approx 1$ \cite{r12}. The density parameter $\Omega_{\phi}$ for the scalar field is given by

\begin{equation} \label{eq.omega}
\Omega_{\phi}=\frac{k^2 \rho_{\phi}}{3 H^2}= x^2 + y^2,
\end{equation}
which is restricted by $\Omega_{b} + \Omega_{\phi}=1$, for a spatially flat universe. In what follows, we shall assume a ``near tracking" situation i.e $\Gamma \approx 1$, which leads to (via equation (~\ref{eq.lambda})) $ \lambda^{\prime} \approx 0$ i.e. $\lambda$ is nearly a constant. The 3-dimensional problem is now effectively reduced to a 2-dimensional autonomous system, with equation (~\ref{eq.x}) and (~\ref{eq.y}).\\
\\
A transformation of the variables to the polar form is now effected with 
\
\\
$x=r \cos\theta$, and $y= r \sin\theta$,
\
so that $r^2=x^2+y^2$ and $\tan {\theta} =\frac{y}{x}$, where $0\leq r \leq \infty$ and $0 \leq \theta \leq 2\pi$. Equation (~\ref{eq.x}) and (~\ref{eq.y}) can now be written in terms of the polar variables,

\begin{equation}\label{eq.r}
r^{\prime} = (\frac{3 \gamma}{2} - 3 \cos^2 \theta) (1-r^2) r,
\end{equation}
\begin{equation}\label{eq.theta}
\theta^{\prime} = (3 \cos \theta - \sqrt{\frac{3}{2}} \lambda r) \sin \theta.
\end{equation}
The fixed points of the system are given by $r^{\prime}=0$ and $\theta^{\prime} =0$. Amongst all the possibilities, the extreme cases like $r^2=0$ (meaning $\Omega _{\phi} =0$) and $r^2=1$ (meaning $\Omega _{\phi}=1$) are excluded. The intention is obviously to have a blend of the quintessence matter and the dark matter. As r is a radial coordinate ranging between 0 and $\infty$, we also exclude the fixed point solution of $ (r, \theta)=(-\frac{\sqrt{3 \gamma}}{\lambda} , \cos^{-1}(-\sqrt{\frac{\gamma}{2}})) $. The only viable option as fixed point is then $ (r, \theta)=(\frac{\sqrt{3 \gamma}}{\lambda} , \cos^{-1}(\sqrt{\frac{\gamma}{2}})) $. In terms of the old variables x and y, this translate into
\begin{equation} \label{eq.fx}
x^2=\frac{3 \gamma^2}{2 \lambda^2},
\end{equation} 
 and
\begin{equation} \label{eq.fy}
y^2= \frac{3 \gamma^2}{2\lambda^2} (1-\frac{\gamma}{2}).
\end{equation}\\
\section{Stability of the solution in polar coordinates:} 
Let us consider the system of equations\\
    $~~~~~~~\dot{x}=f(x,y)$ \\
and $~\dot{y}=g(x,y)$, \\
where an overhead dot denotes differentiation with respect to some parameter ($N$ in the present case).\\
If $u$ and $v$ are the small disturbances from the fixed points  then the system can be linearised in the form 
\\
\
\begin{equation*}
\begin{bmatrix}
\dot{u} \\
\dot{v}
\end{bmatrix} 
= A
\begin{bmatrix}
u \\ 
v
\end{bmatrix}
\end{equation*}

\
\\

 where
 \begin{equation*}
 A=
 \begin{bmatrix}
\dfrac{\partial f}{\partial x} & \dfrac{\partial f}{\partial y} \\
\dfrac{\partial g}{\partial x} & \dfrac{\partial g}{\partial y}
\end{bmatrix} ,
 \end{equation*}
 is called Jacobian matrix at the fixed point. The stability of a fixed point can be determined from the determinant ($ \bigtriangleup $) and trace ($ \tau $) of A at that fixed point. If $ \bigtriangleup < 0$, the eigenvalues are real and have opposite signs hence the fixed point is a saddle point. If $ \bigtriangleup > 0$ and $\tau < 0$ then both the eigenvalues have negative real part hence the fixed point is stable. When $ \bigtriangleup > 0$ and $\tau > 0$ then the fixed point is unstable. Nodes satisfy $\tau^2 - 4 \bigtriangleup > 0$ and spirals satisfy $\tau^2 - 4 \bigtriangleup < 0$ \cite{r19}.
 
 \
 \\
 In the present case, we want to find the stability of the fixed point $ (r, \theta)= (\frac{\sqrt{3 \gamma}}{\lambda} , \cos^{-1}(\sqrt{\frac{\gamma}{2}})) $. Other fixed points are not really of any physical interest as already mentioned. The Jacobian matrix in the present case is \\
\begin{equation*}
A=
\begin{bmatrix}
0 & {3 \sqrt{6} \frac{\gamma}{\lambda} \sqrt{1-\frac{\gamma}{2}}(1-\frac{3 \gamma}{\lambda^2})}\\
-\frac{\sqrt{3}}{2} \lambda \sqrt{1-\frac{\gamma}{2}} & 3 (\gamma -1) - \frac{3 \gamma}{4}
\end{bmatrix} .
 \end{equation*}
For the fixed point, determinant and the trace of the matrix A are, $\bigtriangleup = \frac{9}{\sqrt{2}} \gamma (1- \gamma / 2) (1- \frac{3 \gamma}{\lambda^2})$ and $\tau = 3 (\frac{3 \gamma}{4} -1)$. We consider $\gamma =1$, the matter dominated era with $p=0$. It is known that the fixed point is stable only when $\bigtriangleup > 0$ and $\tau < 0 $. So $\lambda^2 > 3$ is the condition for which the fixed point is stable. To draw the phase portrait of the system, we have plotted  r against $\cos \theta$ instead of that against $\theta$. From the phase plot(figure:1) it is clear that the fixed point is stable in nature which we have also got from the analytical analysis. So any solution of the system around this fixed point will be a stable solution for a wide range of initial values.\\
The phase plot shows that $r=1$ is an invariant submanifold and so is $r=0$. In the second, however, one should note that ($r=0, \cos\theta = 0$) is a saddle type fixed point for our choice of parameters ($\gamma=1 $ and $\lambda = 3 $). It deserves mention that a local subspace bounded by invariant submanifolds given by $r=0$, $r=1$, $\cos\theta = 1$ and $\cos\theta =-1$ has been considered. The reason for restricting to values of $r$ not more than one is the physical requirement that $\Omega_\phi = x^{2}+y^{2} = r^{2}$ indeed lies between 0 and 1.
\begin{figure}[htbp]
\centering
\includegraphics[width=100mm]{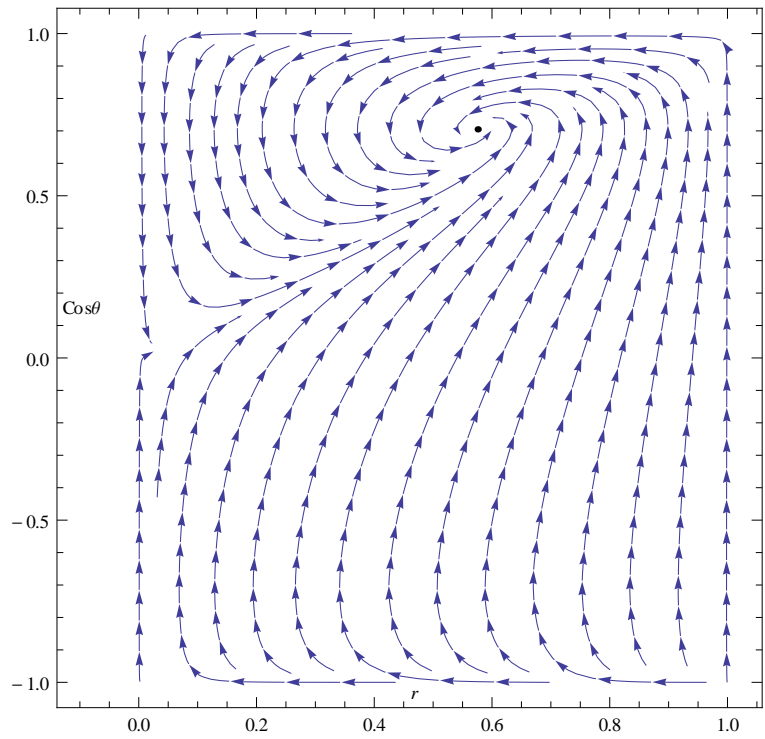}
\caption{Phase plot of the system. The point, P is the fixed point of our interest. The plot clearly shows that the fixed point is stable in nature.The plot is for $\gamma=1$ and $\lambda=3$.}
\end{figure}

 \section{Examples with specific potentials:}  
 \
 \\
 Two examples with specific potentials are now analysed in the present framework. Both the examples are limited to the near tracking zone. One is a hyperbolic potential given by  $V=V_0 \cosh \alpha \phi,$ where $V_0$ and $\alpha$ are constants and $\phi$ is the quintessence scalar field. The other one is an exponential potential, given $ V= e^{\alpha \phi} - \beta$, where $\alpha$ and $\beta$ are positive constants. For a list of quintessence potentials, we refer to the review\cite{r21}.
 \
 \\
 
 \textbf{(i) $V=V_0 \cosh \alpha \phi$}
 \
 \\
 
 The tracking parameter $\Gamma =  V\dfrac{d^2 V}{d \phi^2} / (\dfrac{dV}{d \phi})^2$ in this case is given by $\Gamma = \coth^2 \alpha \phi$. The parameter $\lambda$ is given as $ \lambda = -\frac{1}{k V} \dfrac{dV}{d \phi} = -\frac{\alpha}{k} \tanh \alpha \phi $.
 \\
 
 If the quintessence field is assumed to be a ``tracker" then $\Gamma \approx 1 $. A near tracking condition is assumed given by $\Gamma =1 + \delta $,where $\delta$ is very small. This would imply $\coth^2 \alpha \phi = 1 + \delta$, i.e, scalar field $\alpha \phi$ has a very high value. The parameter $\lambda$ will have a near constant value as for $\delta\longrightarrow 0$, $\tanh \alpha \phi \longrightarrow (1+ \delta)^{-\frac{1}{2}} \approx 1-\frac{\delta}{2}$. So $\lambda$ can be written as $\lambda =- \frac{\alpha}{k} (1-\frac{\delta}{2})$.
 \par At the fixed point, given by equations (~\ref{eq.fx}) and (~\ref{eq.fy}), one has\\
\begin{equation} \label{xcosh}
 x^2=\frac{3 \gamma^2}{2 \lambda^2} = \frac{3 \gamma^2}{2} \frac{k^2}{\alpha^2} \coth^2 \alpha \phi = \frac{3 \gamma^2}{2} \frac{k^2}{\alpha^2} (1+ \delta) ,
\end{equation} \\
   and also 
\begin{equation} \label{ycosh}
y^2 = \frac{3 \gamma}{\lambda^2} (1- \frac{\gamma}{2}).
 \end{equation} \\
    From the definition of $x$ and $y$ (equation(~\ref{eq.trans})) and the form of the potential ($V=V_0 \cosh \alpha \phi$), one can write $\phi$ and H as \\
 \
 \begin{equation} \label{phicosh}
\phi=ln(B a^A),
 \end{equation}
 
 \begin{equation} \label{Hcosh}
 H^2 =\frac{V_0}{3} \frac{\alpha^2}{3 \gamma (1-\frac{\delta}{2}) (1+ \delta)} \frac{B^{\alpha} a^{\alpha A} +B^{-\alpha} a^{-\alpha A}}{2}.
 \end{equation}
 
  Where $A=3(1+ \delta /2) \gamma / \alpha $ and B is arbitary intigration constant.This last equation can be utilised to write the deceleration parameter\\ 
  \begin{equation*}
   q=- \frac{\ddot{a}/a}{\dot{a}^2 / a^2} = - \frac{\dot{H} + H^2}{H^2}
  \end{equation*}
  
   as\\
 \begin{equation} \label{qcosh}
 q =-1-\frac{\alpha A}{2}  \frac{B^{2\alpha } a^{2 \alpha A} -1}{B^{2\alpha } a^{2 \alpha A} +1}
 \end{equation}\\
 
  \begin{figure}[hbtp]
 \includegraphics[scale=0.5]{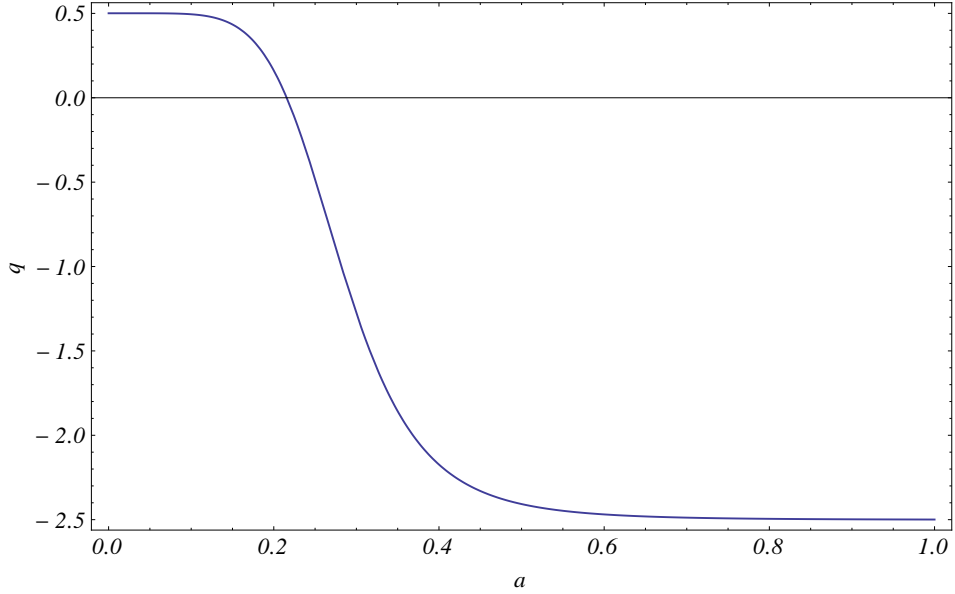} 
\caption{Plot of `q' against `a'.The plot is for the parametric value of $B^{\alpha} = 44.73$, $\delta=0.001$ and $\gamma = 1$.}
 \end{figure}
 
 The plot of `q' against `a' (in units of $a_0$, the present value of a) can now obtained using equation(~\ref{qcosh}). With a tracking condition, $\Gamma \approx 1$, i.e. $\coth^2 \alpha \phi \approx 1$, one finds that `$\phi$' is severely restricted. The tracking condition gives a quadratic equation for the constant $B^{\alpha}$, which may be estimated with the help of the equation (\ref{phicosh}). Such an estimate gives us two values, namely, $B^\alpha = 0.022$ and $B^\alpha = 44.73$. The second solution yields a `q' which at least qualitatively resembles the present acceleration(see figure:2). The problem is that the acceleration sets in quite early in the matter dominated era (near z=4). The other solution $B^\alpha = 0.022$ predicts an acceleration at a distant future and hence not discussed.  \\

 \textbf{(ii) $V=e^{\alpha \phi} - \beta$ }\\
In this case $\Gamma =  V\dfrac{d^2 V}{d \phi^2} / (\dfrac{dV}{d \phi})^2 = \frac{e^{\alpha \phi} - \beta}{e^{\alpha \phi}}$ and $\lambda = -\frac{1}{kV} \dfrac{dV}{d \phi} = -\frac{\alpha e^{\alpha \phi}}{e^{\alpha \phi}-\beta}$. The tracking condition, $\Gamma \approx 1$ demands $\beta \rightarrow 0$. This also would imply $\lambda$ has a near constant value. Near the fixed point $x^2=\frac{3}{2} \frac{\gamma^2}{\lambda^2}$ and $y^2=\frac{3 \gamma^2 (1- \frac{\gamma}{2})}{2 \lambda^2}$, it can be shown that 
\begin{equation} \label{phiexp}
e^{\alpha \phi} - \beta= B a^A
\end{equation}
 and
\begin{equation} \label{Hexp}
H^2=\frac{\alpha^2}{A^2} \frac{\gamma}{1-\gamma/2} \frac{(B a^A + \beta)}{B a^A}.
\end{equation} 

Here B is an integration constant and $A=\frac{3 \gamma}{k}$. Using the expression of $H^2$ one can write
 
 \begin{equation} \label{qexp}
 q=-1-A \frac{B a^A - \beta}{B a^A + \beta}
 \end{equation}
The integration constant B can be estimated from (~\ref{qexp}) by considering present value of q and $a=1$. As $a$ is in units of $a_0$, the present value of $a$, $B=\frac{A - 0.3}{A + 0.3} \beta$. If we take k=1 and $\gamma = 1$ then $A= 3$. Plot of q vs $a$ for this potential agree well the present accelerated expansion of the universe (see figure:3).
\begin{figure}
\includegraphics[scale=0.5]{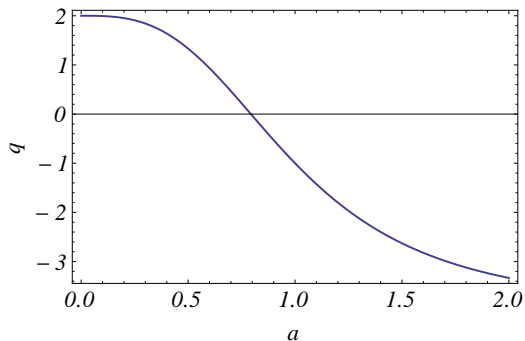} 
\caption{The plot of q vs $a$ for $A=3$, $\gamma=1$ and $\beta=0.00001$.}
\end{figure}\\

\section{Discussion}
The stability of tracking quintessence models for the Universe has been investigated in the present work. With a general tracking condition,the fixed point solutions are not too many. In fact there is one physically relevant generic fixed point, with tracking conditions, leading to a stable solution. The condition for stability, in terms of the fractional rate of change of the quintessence potential, is found out to be $\lambda^2 > 3$ where $ \lambda = - \frac{1}{k V} \frac{dV}{d \phi}$.\\

Two specific potentials, giving rise to the present acceleration, are worked out as examples. It is found that for $V=V_0 \cosh \alpha \phi$, the scalar field is severely restricted to a zone such that $\coth^2 \alpha \phi$ is close to unity. For the other example, $V=e^{\alpha \phi} - \beta$, the conditions do not restrict the scalar field, but rather fine tunes the constants in the potential(eg. $\beta$ must have a very small value).\\

{\bf Acknowledgements:} One of the authors (N.R.) wishes to thank the CSIR (India) for financial support. The authors would like to thank the referee for his comments, particularly on the phase plot, which has improved the quality of the paper.


\begin{thebibliography}{30}

\bibitem{r1} S. Perlmutter et al, Bull. Am. Astron.Soc., {\bf 29}, 1351 (1997).\\
             S. Perlmutter et al, Astrophys. J., {\bf 517}, 565 (1999).\\
             J. L. Tonry et al, Astrophys. J., {\bf 594}, 1 (2003).\\
             S. Bridle, O. Lahav, J.P. Ostriker and P.J. Steihardt, Science, {\bf 299}. 1532 (2003).\\
             G. Hinshaw et al, Astrophys. J. Suppl., {\bf 148}, 135 (2003).\\
             A. Kogut et al, AstroAstrophys. J. Suppl., {\bf 148}, 161 (2003).\\
             D.N. Spergel et al, Astrophys. J. Suppl., {\bf 148}, 175 (2003).\\
             C.L. Bennet at al, Astrophys. J. Suppl., {\bf 148}, 1 (2003).

\bibitem{r20} A.G. Riess et al, Astrophys. J., {\bf 560}, 49 (2001).

\bibitem{r2} T. Padmanabhan and T. Roy Choudury, Mon. Not. R. Astron. Soc., {\bf 344}, 823 (2003).\\
             T. Roy Choudury and T. Padmanabhan, Astron. Astrophys., {\bf 823}, 807 (2005).

\bibitem{r3} V. Sahni and A. Starobinsky, Int. J. Mod. Phys. D, {\bf 9}, 373 (1000).\\
             T. Padmanabhan, Phys. Rep., {\bf 380}, 235 (2003).

\bibitem{r4} J. Martin, astro-ph/0803.4076.

\bibitem{r5} I. Zlatev and P.J. Steinhardt. Phys.Lett.B, {\bf 459}, 570 (1999).

\bibitem{r} N. Banerjee and S. Das, Mod. Phys. Lett. A, {\bf 21}, 2663 (2006).

\bibitem{r6} I. Zlatev, L. Wang and P.J. Steinhardt, Phys. Rev. Lett. {\bf 82}, 896 (1999).\\
             P.J. Steinhardt, L. Wang and I. Zlatev, Phys. Rev.D, {\bf 59}, 123504(1999).\\
             L. Wang, R.R. Caldwell, J.P. Ostriker and P.J. Steinhardt, Astrophys. J., {\bf 530}, 17 (2000).

\bibitem{r7} V.B.Johri Class.Quant.Grav {\bf 19}, 5959 (2002).

\bibitem{r8} L.A.Urena-Lopez and T.Matos Phys.Rev.D, {\bf 62}, 081302 (2000).

\bibitem{r9} M.Sahlen, AR.Liddle, D. Parkinson Phys.Rev.D, {\bf 75}, 023502 (2007).

\bibitem{r10} S.Dodelson, M.Kaplinghat and E. Stewart Phys.Rev.Lett, {\bf 85}, 5276(2000).

\bibitem{r11} P-Y. Wang, C.W Chen and P.Chen. JCAP 2012.


\bibitem{r12} S.C.C.Ng, N.J. Nunes, F.Rosati, Phys.Rev.D, {\bf64} ,083510(2001).

\bibitem{r13} \textit{Dynamical Systems in Cosmology}, J. Wainwright and G.F.R.Ellis (eds); Cambridge University Press, (1997).\\   \textit{Dynamical System and Cosmology}, A.A.Coley, Kluwer Academic Publishers (2003).

\bibitem{r14} L.Lara and M.Castagnins, Int.J.Theor. Phys. {\bf 44}, 1839(2005).

\bibitem{r15} E.Gunzig, V.Faraoni, A.Figeredo, T.M.Rocha Filho and L.Brenig, Class.Quant.Grav., {\bf 17}, 1783(2000).

\bibitem{r16} J.Carot and M.M.collinge, Class.Quanta.Grav., {\bf 20}, 707(2003).

\bibitem{r17} L.Arturo, Urena-Lopez, JCAP, {\bf 0509}, 013(2005).

\bibitem{r18} S.J.Kolitch and B.Hall, arxiv:[gr-qc /9410039].\\
              S.J. Kolitch and D.M. Eardley, Ann. Phys., {\bf 241}, 128, 1995.

\bibitem{r19} \textit{Nonlinear dynamics and chaos: With applications to Physics, Biology, Chemistry and Engineering}, S.H. Strogatz, Westview Press (2001).

\bibitem{r21} V. Sahni, arxiv:{astro-ph/0403324}.
                
\end{thebibliography}
 \end{document}